\documentclass[12pt]{article}
\usepackage{amsmath,amssymb}
\batchmode

\newtheorem{defn}{Definition}
\newtheorem{lem}{Lemma}
\newtheorem{thm}{Theorem}

\newtheorem{cor}{Corollary}

\newcommand{\pr}{\noindent{\bf Proof}. }
\newcommand{\re}{\noindent{\bf Remark}. }
\newcommand{\res}{\noindent{\bf Remarks}. }
\newcommand{\da}{\dagger}
\newcommand{\pa}{\partial}
\newcommand{\one}{\cO(1)}
\newcommand{\half}{{1 \over 2}}

\newcommand{\al}{\alpha}

\newcommand{\de}{\delta}

\newcommand{\Ga}{\Gamma}

\newcommand{\La}{\Lambda}

\newcommand{\Om}{\Omega}
\newcommand{\om}{\omega}

\newcommand{\ep}{\epsilon}
\newcommand{\si}{\sigma}

\newcommand{\cC}{{\cal C}}
\newcommand{\cD}{{\cal D}}

\newcommand{\cO}{{\cal O}}

\newcommand{\cH}{{\cal H}}
\newcommand{\cS}{{\cal S}}

\newcommand{\cF}{{\cal F}}
\newcommand{\cK}{{\cal K}}

\newcommand{\cM}{{\cal M}}

\newcommand{\cL}{{\cal L}}

\newcommand{\cU}{{\cal U}}
\newcommand{\cJ}{{\cal J}}

\newcommand{\bbR}{{\mathbb{R}}}

\newcommand{\bbC}{{\mathbb{C}}}
\newcommand{\bbN}{{\mathbb{N}}}

\begin{document}

\title{ Locality in Free String Field Theory - II.}
\author{ 
J. Dimock\thanks{Research supported by NSF Grant  PHY0070905}\\
Dept. of Mathematics \\
SUNY at Buffalo \\
Buffalo, NY 14214 }
\maketitle
\begin{abstract}
We study the  covariant free bosonic  string field theory and 
explore its locality  (causality) properties.   We find 
covariant string fields which are strictly local and covariant,
 but act on
an unconstrained Hilbert space with an indefinite inner product.
From these we also define observable fields 
which act on the physical Hilbert space with an definite inner product.
These are shown to be  approximately local.
\end{abstract}

\tableofcontents

\newpage

\section{Introduction}

\subsection{Overview}
We continue to explore the locality properties of 
free bosonic string field theory.  The guiding question 
is whether one can find string fields which satisfy the field equation,
are strictly  local  (causal), and are Lorentz covariant. 
Success in this quest  would mean one could define an algebra of local observables of the type
that one usually considers  in quantum field theory  \cite{Haa92}.  We do 
not necessarily expect to succeed in this quest,  but  do expect
that the ways in which we fall short will be of interest.

In an earlier work  \cite{Mar93}, \cite {Low94},  \cite{Dim00}, the problem 
was considered in 
the light cone gauge and   string  fields   were found 
which were local  with respect to the center of mass coordinate.
However these fields were not Lorentz covariant. 

In the present paper we work with a Lorentz covariant formalism
right from the start.   It is the so-called ''old covariant 
quantization''  in which one quantizes first and then imposes the 
constraint.
Before imposing constraints,  we  are able to construct 
string field operators which are    Lorentz covariant
and local in the sense that the commutator of two fields
vanishes when the center of mass coordinates are spacelike separated. 
 However the field operators  act  on  a  Hilbert space  with an indefinite
inner product.  Once the  constraints  are imposed one obtains
 a definite inner product. 
On this space we also define covariant field operators called observable 
fields.
For these observable  fields we establish an approximate locality property.

Our results seem to be  consistent with 
the treatment of   Hata and Oda \cite{HaOd97}
who work  in a BRST formalism.  A general 
account of string field theory can be found in  Thorn  \cite{Tho89}

Another goal of this work is to solidify the mathematical foundations
of covariant string theory.  For earlier work in this direction
see Grundling and Hurst \cite{GrHu93}.

\subsection{Lorentz invariant measures}

We start by  developing some facts about  Lorentz invariant measures on 
the mass shells.  (See also  \cite{ReSi75}).  For any real number  $r$ let 
\begin{equation}
V_r  =  \{  p \in  \bbR^d -\{0\}:  p^2 +r=0 \}
\end{equation}
Here   $p^2  = p \cdot p = \sum_{\mu \nu} \eta_{\mu \nu}p^{\mu}p^{\nu} = -(p^0)^2 + |\vec p|^2$ 
is the Lorentz inner product.  The Lorentz group
is all nonsingular linear transformations preserving $p^2$ and it acts
on $V_r$.

First we 
define a Lorentz invariant volume element on   $V_r$.
Let  $\tau =  dp^0 \wedge dp^1...  \wedge
dp^{d-1}$ be the volume element in   $\bbR^d$.  With $\al(p)=- p^2$  define  $\si$ to be the 
unique   $d-1$ form  on  $V_r$  such that 
\begin{equation}  \label{volume}
d\al \wedge  \si  = \tau 
\end{equation}        
Transforming this by a   proper Lorentz transformation  $\La$ 
we have   $d(\al\circ \La) \wedge  \La^* \si  = \La^* \tau $.
But   $\al  \circ \La =\al$ by definition and $\La^* \tau = \tau$ since  $ \det  \La =1$.
Thus   $d\al \wedge  \La^* \si  = \tau $  and hence  $ \La^* \si = \si$. 

Now   $ \int_{V_r} \ f \ \si$ is
defined for   continuous functions $f$  with  compact support on 
$V_r$.   The map   $ f \to   \int_{V_r} \ f \ \si$ is positive and 
hence there is a positive  measure  $\mu_r$  on  $V_r$  such that 
\begin{equation}   \label{measure}
\int_{V_r}  f \ d \mu_r  =  \int \ f \ \si
\end{equation}
The Lorentz invariance of $\si$ implies the  invariance of   $\mu_r$.

Now we comment on some specific representations of this 
measure, first for   $r \geq 0$.   In this case the hyperboloid has two sheets which 
are   
\begin{equation}
V^{\pm}_r  =  \{  p \in  V_r:   \pm p^0 >0\}
\end{equation}

\begin{lem}  For $r \geq 0$,  let   $f$ have compact support on  $V_r^{\pm}$
and let   $\om_r(p)  = \sqrt {|\vec p|^2  + r}$.  Then
\begin{equation}
\int_{V^{\pm}_r}  f \ d \mu_r 
=  \int f  (\pm \om_r(\vec p) , \vec p)  \frac{d \vec p}{ 2 \om_r(\vec p)} 
\end{equation}
\end{lem}

 \pr For   $V_r^{\pm}$ we can take global coordinates  $\vec p = 
(p^1,...,p^{d-1})$.  In these coordinates we have 
\begin{equation}
\si   =   \frac{1}{ \pa \al / \pa p^0}  dp^1 \wedge ... dp^{d-1}  =
\frac{1}{ 2 p^0}  dp^1 \wedge ... dp^{d-1}
\end{equation}
This is the form   on $V_r^{\pm}$.  Pulling it back to  $\bbR^{d-1}$ with
the inverse coordinate function  $\phi_{\pm}(\vec p )  =  ( \pm \om_r(\vec p) , \vec p)$
we have  
\begin{equation}
\phi^*_{\pm} (\si)  =  \frac{\pm 1}{ 2 \om(\vec p)}  dp^1 \wedge ...\wedge dp^{d-1} 
\end{equation}
Now $\phi_+$ is orientation preserving so 
  $\int_{V^+_r} \ f \ \si$ is evaluated as  $  \int  (f \circ \phi_+) \phi^*_+(\si)$. 
On the other hand  $\phi_-$ is orientation reversing so 
  $\int_{V^-_r} \ f \ \si$ is evaluated as   $-  \int  (f \circ \phi_-) \phi^*_-(\si)$
In either case we get the stated result.
\bigskip

Another representation uses light-cone coordinates and now we allow 
all  $r \in \bbR$.  Light-cone  coordinates  $\hat p = \ell (p)$
are defined  on $\bbR^d$  by     $\hat p =(p^-,  \tilde p, p^+)$  where 
  $p^{\pm} =  (2)^{-1/2}(p^0 \pm  p^{d-1})$ and  $\tilde p = (p^1,...,p^{d-2})$. 
For any function  $f$ let  $\hat f = f \circ \ell^{-1}$ be the expression
in light-cone coordinates. 
Also define  
\begin{equation}
 V_r^{\pm} =  \{ p \in V_r :  \pm p^+ >0 \}
\end{equation}
either in the original coordinates or in light-cone coordinates 
depending on the context.
For   $r \geq 0$ these are again the two components of  $V_r$, but
for $r<0$ they are just open sets in  $V_r$.

\begin{lem}  For any r,   let   $f$ have compact support on  $V_r^{\pm}$.  Then
\begin{equation}  \label{eval2} 
\int_{V^{\pm}_r}  f \ d \mu_r 
=  \int \hat f  \left(\frac{| \tilde p|^2 +r}{2p^+}, \tilde p ,p^+ \right) 
 \frac{ d \tilde  p   dp^+}{ 2 |p^+|} 
\end{equation}
\end{lem}
\bigskip

\pr   The integral $ \int \ f \ \si$  can be evaluated as 
$ \int \ \hat f \ \hat \si$  where   $\hat \si =  ( \ell^{-1})^* \si$.
Since   $ ( \ell^{-1})^* \tau = \tau$  we have   $d \hat  \al \wedge \hat  \si
=\tau$.    On   $V_r^{\pm}$ we can take  coordinates
  $(p^+, \tilde p)$ .   Since   $\tau  = dp^- \wedge dp^1 ...\wedge dp^{d-2} \wedge dp^+$
 and since 
    $\hat    \al(p)=   2 p^+ p^- - |\tilde p |^2 $.
we find that expressed in these coordinates  
\begin{equation}  
\hat \si   =   \frac{1}{ \pa \hat  \al / \pa p^-}  dp^1 \wedge  ... \wedge  dp^{d-2} \wedge dp^+  =
\frac{1}{ 2 p^+}  dp^1 \wedge ...\wedge dp^{d-2} \wedge dp^+
\end{equation}  The inverse coordinate function  
 is  $\phi( p^+, \tilde p)  =  ( (| \tilde p|^2 +r)/2p^+, \tilde p,p^+)$
defined on the half spaces   $\pm p^{+}>0$ in  $\bbR^{d-1}$.
The pull back   $\phi^*(\hat \si)$ to these half spaces   
has the same form.   The function $\phi$ is orientation preserving for $V_r^+$
and orientation reversing for   $V_r^-$.   Thus the integral is
evaluated as
$ \pm   \int  (\hat f \circ \phi) \phi^*(\hat \si)$  respectively, and in either case we get
the stated result.
\bigskip

For   $r  \leq 0$   the 
sets   $V_r^{\pm}$ do not cover all of $V_r$.
However  suppose we define  $p_j^{\pm} =  (2)^{-1/2}(p^0 \pm  p^{j})$  with $j = 1,...,d-1$
and define more sets   $V_{r,j}^{\pm}  =  \{ p \in V_r :  \pm p^+_j >0 \}$ 
and  $\tilde  V_{r,j}^{\pm}  =  \{ p \in V_r :  \pm p^-_j >0 \}$. 
 On each of these sets we can prove a result similar to (\ref{eval2}).
These sets do cover  $V_r$ and by taking a partition of 
unity subordinate to this covering we can express any integral
as a sum of integrals of the type   (\ref{eval2}).

\begin{lem}  \label{decomp} Let  $f$  be a  function on on   $\bbR^d- \{ 0\}$
which is  continuous and has  compact
support.
Let  $f_r$ be the restriction to  $V_r$.  Then
\begin{equation}
\int  f=  \int \left(  \int_{V_r}  f_r d \mu_r \right)  dr
\end{equation}
\end{lem}

\pr 
Let   $U_{r,j}^{\pm}  =  \{ p \in \bbR^d:  \pm p^+_j >0 \}$ 
and  $\tilde  U_{r,j}^{\pm}  =  \{ p \in  \bbR^d :  \pm p^-_j >0 \}$. 
  These cover $\bbR -\{0\}$
and by introducing a subordinate partition of  unity it suffices
to prove the theorem assuming  that $f$ has compact  support in one of these
sets,  for example $U_r^{+}$ defined by  $ p^+  =  p_{d-1}^+>0$.  Then $f_r$ has 
compact support in $V^+_r$ for all $r$.   
We  make the change of variables  
$p^-  \leftrightarrow r$ with  $r = 2p^+p^- - |\tilde p|^2$
and then we have  
\begin{equation}  \label{decompose}
\int \hat  f(p^-, \tilde p, p^+) dp^-  d \tilde p  dp^+
=\int \hat  f(\frac{| \tilde p|^2+r}{2p^+}, \tilde p, p^+)
  \frac{ d \tilde  p  dp^+  dr}{ 2| p^+|} = \int (  \int_{V_r}  f_r d \mu_r)  dr
\end{equation}

\newpage

\section{The single string}

\subsection{Pre-constrained}

We now undertake the covariant quantization of the single string. The  construction 
is mostly standard   \cite{GSW87}.   However, one novelty is that the center of mass  
momentum is treated as a genuine quantum observable with a distribution
of values.  Most treatments take a fixed center of mass momentum. 

For the open string  in $\bbR^d$  the coordinates of the quantum 
string should be  operators  $X^{\mu} (\tau, \sigma)$ define for
 $(\tau, \si)  \in \bbR \times [0, \pi]$ and satisfying
the wave equation
\begin{equation}  \label{waveeqn} 
( \frac {\pa^2 X^{\mu}}{\pa \tau^2} - \frac {\pa^2 X^{\mu}}{\pa \sigma^2} ) =0
\end{equation}
with Neumann boundary conditions  on $[0,\pi]$.
The  operators  $ X^{\mu}$ and the  string momentum  $P^{\mu}= \pa  X^{\mu}/   \pa \tau$
are supposed to satisfy  the equal $\tau$  commutation relations
\begin{equation}   \label{ccr}
[X^{\mu}(\si, \tau ), P^{\nu}(\si', \tau)] = i \pi\de (\si - \si') \eta^{\mu \nu}  
\end{equation}
Corresponding to reparametrization invariance we impose the constraints
\begin{equation}  \label{constraint}
\left(  \frac {\pa X}{\pa \sigma}  \pm   \frac {\pa X}{\pa \tau}  \right)^2 
=0
\end{equation}
That is we ask for states which are annihilated by these  operators.
Finally we ask that the center of mass   $x^{\mu} =
 \pi^{-1}\int_0^{\pi} X^{\mu}(\tau, \sigma) d \sigma$
be parametrized in a forward moving direction.   This means 
we require that the constant center of mass momentum
$p^{\mu}
=dx^{\mu}  /d \tau  =
 \pi^{-1}\int_0^{\pi} P^{\mu}(\tau, \sigma) d \sigma$
should satisfy 
\begin{equation}
  p ^0 = \frac{dx^0}{d \tau} >0
\end{equation}

Classically one can find solutions of the wave 
equation  by expanding in eigenfunctions of the Laplacian
with Neuman boundary conditions, that is in a cosine series.
One finds that 
\begin{equation}  \label{X}
 X^{\mu}(\sigma, \tau ) = x^{\mu} + p^{\mu} \tau   + i\sum_{n \neq 0}\al^{\mu}_n
 e^{-in \tau} \frac{\cos n \sigma }{ n}    
\end{equation}
The quantum operator will be given by the same expression. 
It formally  satisfies  the commutation relations (\ref{ccr})
 if we ask that  $x^{\mu}, p^{\mu} ,\al_n^{\mu}$ be
operators satisfying   the commutation relations
\begin{equation}
\begin{split}
[x^{\mu}, p^{\nu}] = &  i \eta^{\mu \nu}  \\
 [\al_m^{\mu}, \al_n^{\nu}] =& m \de_{m+n}  \eta^{\mu \nu}\label{CCR1}  \\ 
\end{split}
\end{equation} 
 
Here is a construction of these operators.
First   consider the Hilbert space  
$\cL^2(\bbR^d)$.   On this 
space  let    $x^{\mu}$  be  the multiplication operator,  let  
$p_{\nu} = -i \pa/\pa x^{\nu}$  and let  $    p^{\nu}  = \sum_{\nu} \eta^{\mu \nu} p_{\nu}$
 (the spacetime  representation), or else let
$p^{\nu} $ be the multiplication operators and let 
  $x_{\mu} = i \pa /\pa p^{\mu}$, etc.  (the momentum representation).  
In either case these satisfy  (\ref{CCR1}).  A convenient dense  domain 
for these operators is   $\cS(\bbR^d)$,
the Schwartz space of smooth rapidly decreasing functions.  This is invariant 
under the Fourier transform and connects the two representations.

Next let   $\ell^2  = \ell^2( \bbN, \bbC^d)$ be the space of square summable maps 
$f: \bbN \to \bbC^d$.  This has the usual inner product $(f, g)$  and 
 also an  indefinite inner product   $<f, g> $.
They are 
\begin{equation}
\begin{split}  
(f, g )   =& \sum_{n=1}^{\infty} \sum_{\mu }  \overline{f_{\mu n}  }g_{\mu n}\\
<f, g >   =&
 \sum_{n=1}^{\infty}\sum_{\mu \nu}  \eta_{\mu \nu} \overline{f_{\mu n}  }g_{\nu n}\\
\end{split}
\end{equation}
They are related by   $<f, g> =  (f,  J g)$ where  $(Jg)_{\mu}
=  \eta_{\mu \mu } g_{\mu}$. (Thus  $J = \eta$ , but without the geometric 
interpretation).   
Let   $\cF_j(\ell^2)$ be the $j$-fold symmetric tensor product  of   $\ell^2$
and let   $\cF(\ell^2)  = \bigoplus_{j=0}^{\infty}\cF_j(\ell^2)$  be the  bosonic Fock
space over $\ell^2$.   Any unitary operator $U$ on  $\ell^2$ induces a unitary 
$\otimes ^j  U $ on  $\cF_j(\ell^2)$ and hence an operator $\Ga (U)$ on  $\cF(\ell^2)$.
 We define  an indefinite inner product on   $\cF(\ell^2)$  by 
 $<f, g > =(f, \cJ g)$  where   $\cJ = \Ga(J)$.  

For any operator  $\cO$ on  $\cF(\ell^2)$, let  $\cO^*$ be the adjoint
with the definite inner product, and let   $\cO^{\da}$ be the adjoint with
the indefinite inner product.  Hence  $(f, \cO g) = (\cO^*f, g)$ 
 and     $<f, \cO g> = <\cO^{\da }f, g>$ .  They are related by 
$ \cO^*  = \cJ  \cO^{\da}  \cJ  $.

Next define annihilation operators   $a(f), b(f)$ on the n-fold symmetric tensor product by 
\begin{equation}
\begin{split}
a(f) ( f_1 \otimes ... \otimes f_j) = & \sqrt{j} <f,f_1> f_2 \otimes ... \otimes f_j \\
b(f) ( f_1 \otimes ... \otimes f_j) = & \sqrt{j}(f,f_1)\ f_2 \otimes ... \otimes f_j \\
\end{split}
\end{equation}
By restriction these define operators on the symmetric subspace and hence
on  $\cF(\ell^2)$.
   We have   $a(f) = b(J f)$.
The adjoints satisfy  $ a^{\da}(f) = b^*(f)$.  We have  
$[b(f), b^*(g)] = (f,g)$ and 
$[a(f),a^{\da}(g) ] = <f,g>$.

In $\ell^2$ there is a standard basis $e^{\mu}_m$ defined by
$(e^{\mu}_n)_{\nu m} = \de^{\mu}_{ \nu} \de_{m,n}$. 
We  define  for  $n >0$
\begin{equation}
\begin{split}
 \al^{\mu}_n =& \sqrt n \ a(e^{\mu}_n) \\
 \al^{\mu }_{-n} =& \sqrt n \  a^{\da}(e^{\mu}_n)  \\
\end{split}
\end{equation}
These satisfy (\ref{CCR1})  since  $<e^{\mu}_n, e^{\nu}_m> =  \eta^{\mu \nu} \de_{n,m}$.  
 Let  $\cD_0$  be the dense subspace of 
$\cF(\ell^2)$ generated by applying a finite number of  operators  $\al_{-n}^{\mu}$
to the  no excitation state  $\Om_0 =(1,0,0...)$.

Now  consider the Hilbert space  
\begin{equation}  \label{hilbert}
\cL^2(\bbR^d) \otimes \cF(\ell^2))   \approx         \cL^2(\bbR^d, \cF(\ell^2))
\end{equation}
Besides the usual inner product this space has an indefinite inner 
product inherited from  $\cF(\ell^2)$ and defined by 
\begin{equation}
<\psi, \chi>  =  \int  <\psi(p),  \chi(p)> dp
\end{equation}
The operators   $x^{\mu}, p^{\mu}$ in the momentum representation,
and $ \al_n^{\mu}$ all act on this space.
Now we can define the coordinate operator  $X^{\mu}(\si, \tau)$ by (\ref{X}).
It is well defined provided we   interpret it as a distribution in  $\si$ and  
to restrict a nice domain like      $\cS(\bbR^d) \otimes  \cD_0$. It 
does satisfy  (\ref{waveeqn}),(\ref{ccr}).
On the same  domain $(x^{\mu})^{\da }
= x^{\mu}$,  $(p^{\mu})^{\da }= p^{\mu}$,   $(\al^{\mu}_n)^{\da }= \al^{\mu}_{-n}$.
and hence   $(X^{\mu}( \si , \tau))^{\dagger} = X^{\mu}(\si,\tau)$.

We   digress to  discuss representations of the Lorentz group.  First on $\ell^2$ 
there is a representation defined by   $(\La f)_{\mu n} = 
\sum_{\nu}  \La^{\  \nu}_{\mu \  } f_{\nu n}$ 
which preserves the  indefinite inner product.  This induces an
operator  $ \Gamma( \La )$    on  $\cF(\ell^2)$ which also preserves the
indefinite   inner product.
It is not bounded but is at least defined on vectors with a finite number of entries.
Finally for  $a \in \bbR^d$ and  a proper Lorentz transformation  $\La$ 
we define  $U(a, \La)$ on  $ \cL^2(\bbR^d, \cF(\ell^2))$
by 
 \begin{equation}
(U(a, \La)\psi)(p)  = e^{-ip \cdot a}  \Gamma (\La) \psi( \La^{-1} p )
\end{equation}
This is well-defined if  $\psi$ takes values in the domain of  $\Ga(\La)$.
The operators  $U(a, \La)$  give a representation of
the inhomogeneous Lorentz group which preserves the indefinite inner product
since Lebesgue measure is Lorentz invariant.
We note 
that $U(a, \La)^{-1} a(f)  U(a, \La)  =  a (\La^{-1} f)$.   
Since 
  $\La^{-1} e^{\mu}_n =
 \sum_{\nu} (\La^{-1})_{ \nu \ }^{\ \mu} e^{\nu}_n= \sum_{\nu} \La^{ \mu \ }_{\ \nu} e^{\nu}_n$ 
this implies 
\begin{equation}  \label{trans}
U(a, \La)^{-1} \al_n^{\mu}\  U(a, \La)  =
  \sum_{\nu} \La^{ \mu \ }_{\ \nu}   \al^{\nu}_n 
\end{equation}
  We also have  
 $ x^{\mu} \rightarrow
  \sum_{\nu} \La^{ \mu \ }_{\ \nu}  x^{\nu}  + a^{\mu} $,
 and  $p^{\mu}  \rightarrow  \sum_{\nu} \La^{ \mu \ }_{\ \nu}  p^{\nu} $ and  thus  
\begin{equation}
U(a, \La)^{-1} X^{\mu}(\si, \tau)\  U(a, \La)  =
  \sum_{\nu} \La^{ \mu \ }_{\ \nu}   X^{\nu}(\si, \tau)  + a^{\mu}
\end{equation}

Now we turn to the constraint operators  (\ref{constraint}).   
Passing to the Fourier components one finds the operators
\begin{equation}  \label{L}
\begin{split}
 L_0 &= \half \  p^2 + \sum_{n=1}^{\infty}\al_{-n} \cdot \al_n 
 \\
 L_m &= \al_m \cdot p \  + \ \half \sum_{n \neq m,0}\al_{m-n} \cdot \al_n \ \ \ \ \ \  m \neq 0
  \\
\end{split}
\end{equation}
These are well defined on $\cS(\bbR^d) \otimes   \cD_0$ and satisfy 
$L_m^{\da} = L_{-m}$.  

 Instead of asking for states  $\psi$ satisfying 
$L_n \psi = 0$  for all $n$  we make the standard  modification and ask for states 
satisfying 
\begin{equation}
\begin{split} 
(L_0 - 1) \psi =& 0  \\
L_m \psi =& 0 \ \ \ \ \ \ \ \ \ \ \ m>0  \label{first}
\end{split}
\end{equation}

As  usual when quantizing a parametrized theory, the dynamics
are  contained in the constraints.  Suppose we 
 define the operator $M^2$ (not really a square)   on a dense domain 
in   $\cF(\ell^2)$ by 
\begin{equation}
M^2  = 2(N-1)
\end{equation}
where  $N$ is the excitation operator which can be written 
in any of the following forms
\begin{equation}
\begin{split}
 N = & \sum_{n=1}^{\infty}\al_{-n} \cdot \al_n \\
  = & \sum_{n,\mu \nu}  n\ \eta_{\mu \nu}\ a^{\da} (e^{\mu}_n) a (e^{\nu}_n) \\
  = & \sum_{n, \mu}  n\ b^* (e^{\mu}_n) b (e^{\mu}_n) \\ 
\end{split}
\end{equation}
The constraint  $(L_0-1)\psi = 0$ can be written $(p^2 + M^2)\psi = 0$.
In the spacetime representation we have 
\begin{equation} 
\label{KG} (- \square  + M^2 )\psi =0  \end{equation} 
It  the Klein - Gordon equation  for an $\cF(\ell^2)$ valued  function,  and gives the 
evolution in time.  The operator $M^2$  is identified as 
a mass operator.  The next result shows that $M^2$ is self-adjoint 
and has spectrum ${-2,0,2,4,6...}$  with finite multiplicity.

\begin{lem} 
  $N$ is self-adjoint and has  spectrum  $0,1,2,...$  with  finite multiplicity.  
\end{lem}

\pr   Let   $\{ N_{\mu,n} \} $ be a finite sequence  of 
positive integers indexed by  $\mu= 0,1,..., d-1$ and  $n=1,2,...$,
 with at most finitely many  $N_{\mu , n} \neq 0$.
For each such sequence we define a vector 
\begin{equation}
\psi( \{ N_{\mu,n} \} )  =   \prod_{\mu,n}
 \frac{(b^*(e_n^{\mu}))^{ N_{\mu,n}} }{\sqrt{  N_{\mu,n} !}} \Om_0
\end{equation}
This is a orthonormal basis for  $\cF(\ell^2)$ and they are eigenfunctions 
 of $N$ since    
\begin{equation}
   N  \ \psi( \{ N_{\mu,n} \} )  =  (\sum_{\mu,n} n N_{\mu,n} )\  \psi( \{ N_{\mu,n} \} ) 
\end{equation}
This gives the self adjointness and the spectrum.   The finite multiplicity 
follows  since  for any positive integer  $n^*$ there are only a finite number of sequences with 
$\sum_{\mu,n} n N_{\mu,n}  = n^*$.
 
\subsection{Reconfigured}

 The  constraint
(\ref{KG}) cannot be satisfied in   $ \cL^2(\bbR^d, \cF(\ell^2))$.  To impose 
the constraint we will have to modify the Hilbert space.
To begin   we write this space as a direct integral 
over the various mass shells.   For the moment our  purpose
is just to motivate Definition  \ref{defn}
below,  so we  pass over various technicalities such as the exact
definition of the direct integral in this case.  (See however 
Nielsen \cite{Nie80} ).  

  Let   $\| \cdot\|$ denote the norm
in the Fock space  $\cF(\ell^2)$ defined by the definite inner product.   
For    $\psi \in  \cL^2(\bbR^d, \cF(\ell^2))$
the norm squared  can be written as 
\begin{equation}
\int_{\bbR^d} \| \psi (p)\|^2 dp  =
\int_{-\infty}^{\infty} \left( \int_{V_r} \| \psi  ( p)\|^2   d \mu_r(p) \right) dr 
\end{equation} 
This follows from (\ref{decompose}) extended to $\cL^2$ functions. 
Then we can make the identification
\begin{equation}  \label{1}
\cL^2(\bbR^d, \cF(\ell^2), dp ) = 
 \int ^{\oplus} \cL^2(V_r, \cF(\ell^2), d \mu_r)\ dr
\end{equation}
where  
 $\psi  \in \cL^2(\bbR^d, \cF(\ell^2), dp )   $
is identified with the map   $r \to   \psi_r$ ( the restriction of  $\psi$ to $V_r$).
The indefinite inner product on  $\cF(\ell^2)$  induces the same 
on  $ \cL^2(V_r, \cF(\ell^2), d \mu_r)$ and we have,  again by  (\ref{decompose}),      
\begin{equation}  \label{2}
 <\psi , \chi>   =
\int_{-\infty}^{\infty}   <\psi_r, \chi_r> \   dr 
\end{equation} 
Thus the decomposition can be regarded as a decomposition of indefinite inner
product spaces. The operators  $M^2, L_m$ act on  $ \cL^2(V_r, \cF(\ell^2), d \mu_r)$ 
and we have the decompositions 
\begin{equation}  \label{3}
\begin{split}
L_0-1 =&  \int ^{\oplus}\frac12 (-r + M^2)\ dr\\
L_m  = &\int ^{\oplus} L_m \ dr\\
\end{split}
\end{equation}
This means for example that  $(L_m \psi)_r =  L_m \psi_r$.
 Since  the Lorentz group  acts on $V_r$ the  
operators   $U(a,\La )$ act on $ \cL^2(V_r, \cF(\ell^2), d \mu_r)$, and 
they  preserve the indefinite inner product since the measure is
Lorentz invariant. 
Denoting this representation by  $U_r(a , \La)$ we    
 have    
\begin{equation}  \label{4}
U(a, \La) =   \int^{\oplus}  U_r(a, \La)\ dr 
\end{equation}

To impose the constraint we first make a minimal 
reconfiguration of the  
Hilbert space so it can accept the constraints.
Since $M^2$ has  spectrum in $-2,0,2,4,...$  the constraint $(L_0-1)\psi =0$
only has a chance for $r$ in this set. 
Accordingly we pick out these values from the direct integral and form a direct 
sum.  

 At this  stage  we also  impose the 
forward moving condition $p_0 >0$ as much as possible.   
For $r  \geq 0$   we pick out the positive energy part   $V_r^+$
of the mass shell.   This is not possible for   $r<0$  
without losing the representation of  the Lorentz group, and so we leave it alone.

We these ideas in mind we make the following definitions
after  (\ref{1}), (\ref{2}), (\ref{3}), (\ref{4}). 
\begin{defn} \   \label{defn}
\begin{enumerate}
\item
The Hilbert space for the single string is 
\begin{equation}
\begin{split} \label{hrdef}
\cH  =&  \bigoplus_{r= -2,0,2,... } \ \   \cH_r  \\
 \cH_r =& 
\cL^2(   V^{(+)}_r, \cF(\ell^2),  d\mu_r) \\
\end{split}
\end{equation}
where   $  V^{(+)}_r = V_r^+$ for  $r \geq 0$ and   $  V^{(+)}_r = V_r$ for  $r < 0$ .
\item For   
$\psi, \chi$ in  $\cH$ with  components  $\psi_r, \chi_r$ 
 an indefinite inner product is defined by  
\begin{equation}  
 <\psi , \chi>   =
 \sum_{r }  <\psi_r, \chi_r>     
\end{equation} 
\item A representation of the inhomogeneous Lorentz group 
is defined by 
\begin{equation}
U(a, \La) =  \bigoplus_r  U_r(a, \La)  
\end{equation}
\item The constraint operators are defined by   
\begin{equation}
\begin{split}
L_0-1 = & \bigoplus_r  \frac12 (-r + M^2) \\
L_m  = & \bigoplus_r L_m \\
\end{split}
\end{equation}
\end{enumerate}
\end{defn}

Let us be more precise about the domains of 
$N, M^2, L_m$.  We will define them  as closed operators on  $\cH_r$ and then 
the above equations  define them as closed operators
on  $\cH$.  As we have noted $N$ (or $M^2$) is 
self-adjoint on  $\cF(\ell^2)$ and we define  $N$ on   $\cH_r$ by
$(N\psi)(p) = N \psi(p)$ with domain
\begin{equation} 
D(N)  =  \{  \psi \in \cH_r :  \psi(p) \in  D(N) \ a.e.\  p \ , \  
\int \| N\psi(p) \|^2 d\mu_r(p) \  < \infty \} 
\end{equation}
Similarly for each  $p$ let  $L_m(p)$ be the closure of the operator (\ref{L})
 defined on  $\cD_0 \subset \cF(\ell^2)$. 
(It is closable since the adjoint
   $L_m^*(p)  =  \cJ L_{-m}  \cJ $  is  densely defined).
  Then  define  $L_m$ on  $\cH_r$
by   $ (L_m \psi)(p) = L_m(p) \psi(p)$ with domain
\begin{equation} 
D(L_m)  =  \{  \psi \in \cH_r :  \psi(p) \in  D(L_m(p))  \ a.e.\  p \  , \  
\int \| L_m(p) \psi(p) \|^2 d\mu_r(p) \  < \infty \} 
\end{equation}
With these domains we  have:

\begin{lem} 
 $N$ and  $M^2$ are  self adjoint,   and  $L_m$ is closed.
\end{lem}

\pr  Start with the second.   Let   $\psi_j \in D(L_m)$  satisfy   $\psi_j \rightarrow  \psi$ and $ L_m \psi_j
\rightarrow \chi$.    Then there exist a subsequence such that for almost every $p$ we
have    $\psi_{j_k}(p) \rightarrow  \psi(p)$  and   $L_m(p)\psi_{j_k}(p) \rightarrow  \chi(p)$.
(This is a corollary of the proof that  vector-valued $\cL^p$ spaces are complete
\cite{HiPh57}).
Since  $L_m(p)$ is closed it follows that   $\psi(p) \in  D(L_m(p))$ and    
$L_m(p) \psi(p) = \chi(p)$ for  a.e. $p$.     
Hence   $\psi \in D(L_m)$  and  $L_m \psi = \chi$.  Thus $L_m$ is closed

The same argument shows that   $N$ is closed.   It is also symmetric and 
since it has a dense set of analytic vectors,  for example  $\cC^{\infty}_0 \otimes \cD_0$,
it is self-adjoint.

\subsection{Constrained}

Now let  
$\cH'$ be the subspace of  $\cH$ satisfying  the constraints  (\ref{first}).
We have   
\begin{equation}     \label{prime}
\begin{split}
 \cH' =& \bigoplus_{r= -2,0,2,... }  \cH_r'   \\
\cH_r'  =  & \{  \psi \in \cH_r: (-r + M^2)\psi =0,\  L_m \psi = 0 \ \rm{for} \ m>0 \}  \\
\end{split}
\end{equation}
 Note that a function  $\psi \in \cH_{-2}$ is   in  $\cH'_{-2}$  iff both $N\psi=0$
and $L_m\psi=0$ which is true iff  $\psi$ takes values in $\cF_0(\ell^2)  \approx \bbC$.
 Thus $\cH'_{-2}  =   L^2(   V_{-2}, \cF_0(\ell^2),  d\mu_{-2})$.  These are
 the \textit{tachyons}.

Next  we consider  the \textit{isotropic}  or \textit{spurious} elements 
in    $\cH'$  which are defined 
by   
$\cH'' = \cH' \cap (\cH')^{\perp}$.  Here the  orthogonal subspace is defined by 
the indefinite inner product.
Vectors in  $\cH''$ satisfy   $< \psi, \psi > = 0$.
The subspace has the form
\begin{equation}
\begin{split}
\cH'' =   &      \bigoplus_r   \cH''_r  \\
\cH''_r = &   \cH_r' \cap (\cH_r')^{\perp}\\
\end{split}
\end{equation}
Now  let  $ \cH^{phys} = \cH' / \cH'' $.   We  identify 
\begin{equation}
\begin{split}
 \cH^{phys} = &     \bigoplus_r   \cH^{phys}_r  \\
 \cH^{phys}_r =& \cH'_r / \cH''_r  \\
\end{split}
\end{equation}
The indefinite  inner product  on $\cH'$
lifts to   $ \cH^{phys} $  and is the direct sum of the 
the inner products on  $\cH_r^{phys}$ lifted from  $\cH_r'$.

Now we establish  the famous  no-ghost theorem.  Our method is to 
reduce the result to a statement pointwise in $p$ and them
quote the literature.

\begin{lem}  For $d=26$,
$<.,.>$ is positive definite on   $\cH^{phys}_r$ and  $\cH^{phys}$.
\end{lem}

\pr   It suffices to prove the result   on $\cH^{phys}_r$. 
The space  $ \cH_r  = \cL^2(   V_r^{(+)}, \cF(\ell^2),  d\mu_r)$
 can be regarded as a constant fiber direct integral
\begin{equation}
\cH_r  = 
  \int^{\oplus}_{V^{(+)}_r} \cF(\ell^2) d \mu_r(p)
\end{equation}
Indeed the former can be taken as the definition of the latter,  so this just 
amounts to a change in notation.
We have the decompositions $(-r+M^2) = \int^{\oplus}(-r+ M^2) d \mu_r(p)$
and $L_m = \int^{\oplus}  L_m(p) d \mu_r(p)$. The constrained space can 
be characterized as 
\begin{equation}
\begin{split}
\cH'_r  =  &
  \int^{\oplus}_{V^{(+)}_r}  \cH'(p) d \mu_r(p) \\
\cH'(p)  =  & \{  \psi \in \cF(\ell^2): (p^2 + M^2)\psi =0,  L_m(p) \psi = 0 \ \rm{for} \ m>0 \}  \\
\end{split}
\end{equation}
This means  $\psi \in \cH_r'$ iff  $\psi(p)  \in  \cH'(p)$ for a.e. $p \in V^{(+)}_r$.
 Since eigenvalues of  $M^2$ have finite multiplicity,
 $\cH'(p)$ is  finite dimensional.  Note  also  $\cH'(p)  \subset \cD_0$.

Next we have    
$(\cH'_r )^{\perp} =  
  \int^{\oplus}(\cH'(p))^{\perp} d \mu_r(p) $ and it follows that   
\begin{equation}
\begin{split}
\cH''_r  =  &
  \int^{\oplus}_{V^{(+)}_r}  \cH''(p) d \mu_r(p) \\
\cH''(p) =&  \cH'(p) \cap  \cH'(p)^{\perp} \\
\end{split}
\end{equation}
Now for  $\psi, \chi \in \cH_r'$ we have   $\psi- \chi \in \cH_r''$
if and only if $\psi(p), \chi(p) \in \cH'(p)$ satisfy  
 $\psi(p)- \chi(p) \in \cH''(p)$ for a.e. $p$.  Thus equivalence classes 
can be defined pointwise which we write as
\begin{equation}
\begin{split}
\cH^{phys}_r  = & 
  \int^{\oplus}_{V^{(+)}_r}  \cH^{phys}(p) d \mu_r(p)\\
\cH^{phys} (p)  =& \cH'(p) /\cH''(p)  \\
\end{split}
\end{equation}

Now it suffices to prove that the inner product is positive 
definite on     $\cH_{phys} (p)$.   A proof of this can
be found in  Frenkel, Garland, and Zuckerman
\cite{FGZ86}.  They also compute the dimension of this space 
and show it depends only on  $p^2$.  For the original proofs see Brower \cite{Bro72} and 
Goddard and Thorn \cite{GoTh72}.

\begin{lem}   $U(a, \La)$ determines a  unitary representation of 
the inhomogeneous Lorentz group on  $\cH^{phys}$.   
\end{lem}

\pr   $U_r(a, \La)$ is defined on all of  $\cH_r'$.
It   preserves  $\cH'_r$  since 
$[L_m, U_r(a, \La)] =  0$ by(\ref{trans}).  
  If  $\chi \in (\cH_r')^{\perp}$ then
  for   $\psi \in  \cH_r'$
 we have 
$< \psi,  U_r(a, \La)\chi > = < U_r(a,\La)^{-1} \psi,\chi >=0$
and so  $ U_r(a, \La)\chi \in (\cH_r')^{\perp}$.
Thus  $ U_r(a, \La)$ preserves  $\cH''_r$ and so it lifts
to  $\cH^{phys}_r$.   Since it is still inner product preserving
it is unitary.

\subsection{Physical States}

We want to exhibit some non-trivial smooth elements of 
$\cH'_r \subset \cL^2(V^{(+)}_r,\cF(\ell^2), d \mu_r)$.
These then determine well-behaved vectors in  $\cH_r^{phys}$ and  $\cH^{phys}$.  We start
with  the fact  the for  each  $p$ the spaces  $\cH'(p)$ are non-trivial (finite dimensional )
vector spaces.

\begin{lem}  \label{lemma}
$\Ga(\La)$ is a bijection  from  $\cH'( p)$
to   $\cH'(\La p)$ .
\end{lem}

\pr 
First note that  $\Ga(\La)$ is defined on  $\cH'(p)$
since $\cH'(p)  \subset \cD_0$.
Then we use    (\ref{trans})
to conclude that  $\Ga( \La)^{-1} M^2\  \Ga( \La)  =M^2 $ and that
$\Ga( \La)^{-1} L_m(p)\  \Ga( \La)  =
   L_m(\La^{-1} p) $.
This gives the result.  

\begin{lem}
 For any  $q \in  V^{(+)}_r $  there is a neighborhood
$U$ of $q$  in  $V^{(+)}_r $ and a smooth  family   $\La(p,q)$ of Lorentz transformations
defined for $p \in  U$ such that   $\La(p,q)  \  q = p$.
\end{lem}

\pr   The proof uses  the following well-known  fact  (see for example \cite{War83}).   Let 
$G$ be a Lie group with closed subgroup  $H$.  Let  $\pi: G \rightarrow  G/H $ be 
the projection onto the coset space.  Then there are local smooth 
sections.  That is for any point in $G/H$  there is a  neighborhood $U$
and a smooth  map   $s: U \rightarrow  G$ such that   $\pi \circ s = id $.

In our case 
let  $\cL$ be the proper Lorentz transformations ($\cL$ = a component of $SO(d-1,1)$).
Fix $q$ and let   $H$ be the subgroup
which leaves  $q$ invariant.   (e.g. if $r>0$ then  $H = SO(d-1)$).
Since  $\cL$ acts transitively on  $V^{(+)}_r$ we have that the map 
$\La \to \La q$ from   $\cL $ to $V^{(+)}_r$  lifts to a diffeomorphism
$\La H  \to  \La q$ from  the coset
space  $\cL/H$ to  $V^{(+)}_r$.  This identifies the 
two spaces.

Now from the general result there is a neighborhood   $U$ of $H$  in  $\cL/H$
and a map   $s:  U \to  \cL$  such that   
$s(\La H ) H = \La H$ for all  $\La H  \subset U$.  Equivalently 
we can regard $U$ as a neighborhood of $q$ in $V^{(+)}_r$ and 
have a map  $s:  U \to  \cL$ satisfying   $s(p)q =p$ for all $p \in U$.  Defining  $\La(p,q)
= s(p)$ we have the result.
\bigskip

\begin{lem}  Let  $q, U$,  and $\La(p,q)$ be as above.  \label{states}
\begin{enumerate}
\item   $\Ga(\La(p,q))$ is a bijection from $\cH'(q)$
to  $\cH'(p)$.
\item  There  exist  $\psi \in \cC^{\infty}(U, \cF(\ell^2))$
such that   $\psi(p)  \in \cH'(p)$ for all $p \in U$.
\item  There  exist  $\psi \in \cC^{\infty}_0(V^{(+)}_r, \cF(\ell^2))$
such that   $\psi(p)  \in \cH'(p)$ for all $p \in V_r$, i.e.  $\psi \in \cH_r'$.
\end{enumerate}
\end{lem}

\pr  The first follows from lemma \ref{lemma}.    For the second let
  $\psi_0 \in \cC^{\infty}_0( U , \cH'(q))$ and take $\psi(p) = \Ga( \La(p,q)) \psi_0(p)$.
 Multiplying by  
 $\chi \in \cC^{\infty}_0(U)$ gives a functions satisfying the third condition.
By adding functions for different neighborhoods $U$ we get a rich class
of functions. \bigskip

\newpage
\res
\begin{enumerate}
\item  There is another way to look at  this result.  Let  $\xi$ the set of all
pairs 
  $\{(p,\psi) \}$ where   $p \in  V_r$  and  $\psi \in  \cH'(p)$,
and let $\xi_U$ be the subset of pairs with  $p \in U$.
 We have  smooth  mappings   $(p , \psi) \to (p,\Ga( \La(p,q)^{-1} ) \psi)$     
from  $\xi_U$  to   $U \times \cH'(q)$ and hence from  $\xi_U$ to 
 $U \times \bbC^s$ where $s$ is the dimension of  $\cH'(q)$.
Covering $V_r^{(+)}$ by such charts gives 
 $\xi$ the structure of a  smooth  vector bundle.
Such bundles have  smooth sections which our result.

\item   If   $r>0$  then the neighborhood $U$ can be 
taken to be all  of  $V^+_r$.    In this case we can take  $q = (\sqrt{r},0,0,..)$
and let  $\La(p,q)$ be the standard boost to $p$.  ( See  Weinberg \cite{Wei95},
equation (2.5.24) for the explicit formula).

\end{enumerate}

\section{String field theory}

We  develop the string field theory by 
taking the dynamical equation satisfied by the 
 single string  wave equation,
specializing to real solutions, treating these solutions as 
 a  classical Hamiltonian system, quantizing this 
system, and then finally imposing the constraints.
This is ''second quantization'' , and the exact status
of this process  is 
always a little ambiguous \ldots  which quantization was  the genuine  quantization?
Whatever attitude one takes one ends up at the same place.  
In any case  the  quantization  process is just meant to be suggestive
of a true quantum theory.

 Our formulation of the problem uses techniques which have been useful 
in the study of  quantum field theory in curved spacetime   
\cite{Ish78}, \cite{Dim80}, \cite{Dim92}. 

\subsection{String field  equation}

We start by defining real elements of  $\cF(\ell^2)$.  These are vectors
satisfying  $C \psi = \psi$ where  $C$ is some conjugation on  $\cF(\ell^2)$.
A conjugation is an anti-linear isometry satisfying  $C^2 = 1$.  We
also want our conjugation to satisfy $[C,M^2]=0$ and $[C,\cJ]=0$.
Then also   $\overline{<\psi, \chi>  }= <C \psi, C \chi>$.
For example one could take $ C_0 = \Ga(c_0)$ where $c_0$ is 
the usual conjugation  $c_0 \psi = \bar \psi $ on $\ell^2$.
In the following we just suppose that some  $C$ satisfying the
above conditions has been chosen.

Now we study  the Klein-Gordon equation:
\begin{equation}
(- \square  + M^2) U=0
\end{equation}
for  functions  $U: \bbR^d \to \cF(\ell^2)  $.
Given real  $F_0,G_0 \in  \cC^{\infty}_0(\bbR^{d-1}, \cF(\ell^2))$
there is a unique smooth real solution  $U$ such that  $U=F_0$
and $\pa U/ \pa x^0 = G_0$ on some surface  $x^0 = t_0$, called 
a Cauchy surface.
The solution has compact support on any other Cauchy surface  $x^0 =t$.
   Such solutions will be called 
\textit{regular}.

Associated with this  equation there is a real bilinear form. 
For any functions  
$U(x) = U(x^0, \vec x)$ and  $V(x) = V(x^0, \vec x)$  it is defined by   
\begin{equation}
\sigma_t (U,V)  =  \int_{x_0 = t }( <U(x), \frac{\pa  V}{ \pa x^0}(x) >-
  < \frac{\pa U}{ \pa x^0}(x)  ,V(x)>)\ d \vec x
\end{equation}
Green's identity states that for $t>s$ 
\begin{equation} 
\begin{split} 
&\sigma_t (U,V)-\sigma_s (U,V)  =     \\
&\ \ \ \ \ \  \int_{s < x^0 < t} 
(<U(x) ,((-\square +M^2) V)(x) >
- <((-\square +M^2) U)(x),V(x)> ) 
 dx \\
\end{split}
\end{equation}
Thus if   $U,V$ are  regular solutions  then  $\sigma_t(U,V)$ is independent of 
$t$ and is just denoted   $\sigma(U,V)$. 
This form is skew symmetric and  non-degenerate on the space of 
regular solutions, i.e. it is symplectic.   This symplectic form will be 
the basis on quantization.

 But first we develop some facts about fundamental solutions
for our   KG equation.   The following results are standard
for the scalar KG equation.
Advanced and retarded fundamental solutions $E^{\pm}$ are 
operators on  functions 
$F \in \cC^{\infty}_0(\bbR^d  ,   \cF(\ell^2))$  defined by 
\begin{equation}
(E^{\pm} F)(x)  = \frac{1}{(2 \pi)^{d/2}}  \int_{\Ga_{\pm} \times \bbR^{d-1}}
 \frac{ e^{ip\cdot x}}{p^2 + M^2} \ 
\tilde  F(p) dp 
\end{equation}
The  $p^0$ contour $\Gamma_{\pm} $ is the real line shifted slightly above/below
the  real axis.   The exact choice does not matter because  $\tilde F$ is entire
and rapidly decreasing in real directions. 
Also for $x \in\bbR^d$ let   $J^{\pm}(x) = \{ y:  (x-y)^2 <0, \pm(x^0-y^0) >0 \} $
be the
past  or future of  $x$.  For  $A \subset \bbR^d$   
define   $J^{\pm} (A) =  \cup_{x \in A}  J^{\pm}(x)$.

\begin{lem}  For  $F \in
\cC^{\infty}_0(\bbR^d  ,  \cF(\ell^2))$ we have 
$(-\square + M^2) E^{\pm} F=F$ and   $supp(E^{\pm} F) \subset J^{\pm} (suppF)$.
\end{lem}

\pr 
Let  $P_r$ be the  operator on  $\cF(\ell^2)$  which is the 
projection onto the eigenspace  $M^2 = r$.
On the range of  $P_r$ we have   $E^{\pm}F = E^{\pm}_r F$ where 
\begin{equation}
(E_r^{\pm} F)(x)  =\frac{1}{ (2 \pi)^{d/2}}  \int_{\Ga_{\pm} \times \bbR^{d-1}}
 \frac{ e^{ip\cdot x}}{p^2 + r}
\ \tilde  F(p) dp 
\end{equation}
These are  the advanced/retarded fundamental solutions 
for the Klein-Gordon equation with mass $r$ and they  satisfy  $(-\square + r)E_r^{\pm}F=F$ 
and  $ supp (E_r^{\pm}  F) \subset  J^{\pm}( supp F)$.

Now $ E^{\pm}$ are fundamental solutions since  $P_r (- \square  + M^2 ) E^{\pm}F
=  (- \square  +  r ) E^{\pm}_r P_rF =P_rF$.  For the support property we have
 $supp(E^{\pm} F)  \subset \cup_r supp (P_r E^{\pm} F)$ .
But $P_r E^{\pm} F = E_r^{\pm}P_r F $ and  
$supp(  E_r^{\pm}P_r F) \subset J^{\pm} (supp (P_rF))  \subset J^{\pm}(supp F)$
and hence the result.
\bigskip

 The propagator function is defined by  $E =  E^+ - E^-$.
Then $U = EF$ is a regular solution.  In fact we have:
\begin{lem} \
\begin{enumerate}
\item   $U$ is a regular solution of  $(- \square  + M^2)U =0$ iff it can be written  $U = EF$
with $F \in
\cC^{\infty}_0(\bbR^d  ,  \cF(\ell^2))$
\item  $F \in \cC^{\infty}_0(\bbR^d  ,   \cF(\ell^2))$ satisfies  $EF=0$  iff
$F= (- \square  + M^2)H$ for some 
$H \in \cC^{\infty}_0(\bbR^d  ,   \cF(\ell^2))$
\end{enumerate}
\end{lem}

\pr   Let $U$ be a regular solution and let   $ \theta = \theta (x^0)$
 be smooth and satisfy  $\theta = 1 $ for $x^0$ sufficiently positive
 and  $\theta = 0$ for $x^0$ sufficiently negative.  Define 
 $F=(- \square +M^2) ( \theta U) = -(-\square + M^2) (( 1- \theta ) U)$.
 Then $F$  has compact
support. Next note that   $ \theta U = E^+ F$ since the difference satisfies the $KG$
equation and vanishes in the distant past. 
Similarly    $- (( 1- \theta ) U)  = E^-F$. 
Taking the difference of the last two
equations gives $U=EF$ .  This proves the first claim. 

For the second suppose    $F  = (- \square  + M^2 )H$.
Then   $EF= (- \square  + M^2 )EH =0$. On the other hand if   $EF=0$, then    $H = E^{\pm}F $
has compact  support and   $ (- \square  + M^2 )H = F$. 
\bigskip

 The next  identity establishes a connection
between any solution and its
values on any Cauchy surface  (all in the sense of distributions). 
We define 
\begin{equation} 
<U,F>  =  \int <U(x), F(x)>  dx  
\end{equation}

\begin{lem}  For  $F \in \cC^{\infty}_0(\bbR^d  ,   \cF(\ell^2))$
and any regular solution  $U$ :
\begin{equation}  \label{connection}
\sigma(U,EF)  = <U,F> 
\end{equation}
or equivalently   For  $F,G \in \cC^{\infty}_0(\bbR^d  ,   \cF(\ell^2))$  
\begin{equation}  \label{connection2}
\sigma(EF,EG)  = <EF,G>  = -<F,EG> 
\end{equation}
\end{lem}

\pr  By Green's   identity  we have  for  $t<0$   
\begin{equation}
\si_0 (U,E^+ F) - \si_t (U,E^+ F)  =  \int_{t <x^0<0}
<U(x), F(x)> dx
\end{equation} 
Letting   $t \to -\infty$  we get an expression for $\si_0 (U,E^+ F) $.  
Similarly we get an expression  for $\si_0 (U,E^- F) $ .  They
are   
\begin{equation}
\sigma_0(U,E^{\pm}f)  = \pm  \int_{ \mp x^0 >0} <U(x),F(x)> dx 
\end{equation} 
Take the difference to obtain the result.  
\bigskip

We next want to make a connection with the single string Hilbert space  
$\cH=\oplus_r \cH_r$.
Given  $F \in \cC^{\infty}_0(\bbR^d  ,   \cF(\ell^2))$
 we define   $\Pi F  \in \cH$ 
 by specifying that $(\Pi F)_r \in  \cH_r$  is 
obtained by   taking the Fourier transform, projecting onto the 
subspace  $M^2=r$ with $P_r$, and then restricting to $V^{(+)}_r$.  
More precisely for    $p \in  V^{(+)}_r$ we define    
\begin{equation}
(\Pi F)_r(p)    =  \sqrt{ 2 \pi} \ P_r \tilde F ( p )  
\end{equation}
We will need to exclude tachyons, so we restrict to 
 functions  $F$ which take values in 
\begin{equation}
\cF_+ (\ell^2) \equiv  (\cF_0(\ell^2))^{\perp}  \equiv   \bigoplus_{j \geq 1}  \cF_j(\ell^2) 
\end{equation}
We have  $M^2 \geq 0$ on this subspace.  Hence for such $F$ , $(\Pi F)_{-2} =0$  and 
hence $\Pi F  \in \cH_+$ the no-tachyon subspace of  $\cH$: 
\begin{equation}
\cH_+   = \bigoplus_{r \geq 0}  \cH_r
\end{equation}

\begin{lem}    For (real) $F,G \in \cC^{\infty}_0(\bbR^d  ,   \cF_+(\ell^2))$
\begin{equation}  \label{basiceq}
   \sigma(EF,EG)  =  <E F,G > = 2 {\rm Im} <\Pi F, \Pi  G> 
\end{equation}
\end{lem} 
\bigskip

\pr  Only the second identity needs proof.
 We compute  with  $E_r = E_r^+ - E_r^-$  
\begin{equation}
\begin{split}
<EF,G>=& \sum_{r \geq 0}< E_rP_rF, P_r G >\\
 =&  \sum_{r\geq 0}2\ {\rm Im} <(\Pi F)_r, (\Pi G)_r> 
= 2 \ {\rm Im} <\Pi F, \Pi  G>  \\  
\end{split}
\end{equation}
The second step follows since  for $r \geq 0 $ and any   $F,G$
\begin{equation}
\begin{split}
   <E_r F, G>  
 = & -\int_{\Gamma_+ - \Gamma_-}  
< \tilde F( \bar p) , \tilde G(p)>\frac {1} { p^2 +r} dp  \\  
= &-2 \pi i\{\int < \tilde  F(\om_r(\vec p), \vec p)   ,
  \tilde G(\om_r(\vec p),\vec p)>
\frac {  d \vec p }{2\om_r(\vec p)} -c.c. \} \\ 
= & \  2\ {\rm Im} < \sqrt{2\pi} \tilde F | V_r^+, \sqrt{2\pi} \tilde G | V_r^+>     \\
\end{split}
\end{equation} 
In the second step we have evaluated the contour integral 
by taking residues at   $p^0 = \pm  \om (\vec p)$.  We have also 
used  
  $ \overline{< \tilde F(p) , \tilde G(p)>} =<\cC \tilde F(p) ,\cC\tilde G(p)>
 = < \tilde F(-p) , \tilde G(-p)>$ for $p$ real  , a consequence of the reality of 
$F,G$.

\subsection{String field operator}

Now we quantize solutions of the string field equation.
We take as our phase space the space of all regular solutions $\Phi$
of    $(-\square + M^2) \Phi =0$  with  symplectic form 
  $\sigma( \Phi, \Phi')$ defined
previously.  For each solution $U$ there is a  function   $\Phi \to 
\sigma(\Phi, U)$ on the phase space.   
We quantize these functions by  replacing them by operators on a complex Hilbert
space, also denoted  $\sigma(\Phi, U)$,
which are required to satisfy
\begin{equation}  \label{comm1}
[\sigma (\Phi,  U), \sigma (\Phi, V)] = i \sigma (U, V) 
\end{equation}
This  looks more familiar if we  identify solutions with 
their data on some Cauchy surface.   Then the operators
are   $\si(\Phi_0,\Pi_0;F_0,G_0)  =  \Phi_0(G_0) - \Pi_0(F_0)$
 and the commutator is written  
$ [\sigma (\Phi_0, \Pi_0; F_0,G_0), \sigma (\Phi_0, \Pi_0; F'_0,G'_0)]
 = i \sigma (F_0,G_0;F'_0,G'_0)$.  As a special case  we have the standard   
$ [ \Phi_0(G_0), \Pi_0(F_0)] = i<G_0,F_0>$.

The full spacetime field operator is obtained from the operators
$\sigma(\Phi, U)$ just as in the classical case.   
Following   (\ref{connection})  we define  the field operator
as a distribution by   
\begin{equation}   \label{connection3}
\Phi (F)   =  \sigma ( \Phi,EF) 
\end{equation}
   Then   $\Phi(F)$  satisfies the field equation and has 
a local commutator as the next result shows.

\begin{lem}   let  $\sigma (\Phi,  U)$ be a family of operators  indexed
by  regular solutions  $U$ of the KG equation and satisfying   
(\ref{comm1}).  Then the operators  $\Phi (F)  =  \sigma ( \Phi,EF) $  defined
for   $F   \in  \cC^{\infty}_0( \bbR^d,  \cF(\ell^2))$  
satisfies
\begin{equation}  
\begin{split}
\Phi( (-\square + M^2 )F) =  &   0 \\ 
[ \Phi (F), \Phi(G)]  =& -i  < F,EG > \\
 \label{basic}
\end{split}
\end{equation}
Furthermore every operator valued distribution   $\Phi(F)$ 
satisfying (\ref{basic})  arises in this way.
\end{lem}

\pr  The field equation 
 follows from $E (- \square + M^2) F =0$
and the commutator  follows from  the identity (\ref{connection2}).
For the converse given  $\Phi(F)$
we    define   $ \sigma ( \Phi,U)  = \Phi(F)$  
for any $F$ such that   $U = EF$.  
To see that this is well defined we have to show that   if $EF_1 = EF_2$
then  $\Phi(F_1) = \Phi(F_2)$,  or if  $EF=0$ then $\Phi(F) = 0$.
But we have seen that $EF=0$ implies  $F = (-\square + M^2) H$ and hence the result follows.
 The operators $ \sigma ( \Phi,U)$ have the commutator (\ref{comm1})  
again by the 
identity  (\ref{connection2}).
\bigskip

\res
\begin{enumerate}
\item
Since   $\sigma (U, V) $ is a symplectic form, representations of
 (\ref{comm1}) do exist on general principles.
 Thus string field theories exist.   Furthermore the spacetime field  $\Phi(F)$
defined by (\ref{connection3}) is strictly local because  if $supp ( F)$ and $supp (G)$
are spacelike separated,  then  $supp (EF)$ and $supp (G)$  do not overlap
and hence   $[\Phi(F), \Phi(G)] = -i<F,EG> =0$.
All this holds without suppressing the negative mass part of the equation!
 \item
However this  is not the end of the story.   We actually 
want the particular  representation in which time translation is unitarily implemented
with positive energy.  (One can think of this as the forward moving condition
again).  Choosing a particular representation requires  a complex structure 
or a "one-particle structure" on phase space.  
These are  equivalent  to  expressing the symplectic form  $\sigma(EF,EG)$ as the 
imaginary part of an inner product on some complex Hilbert
space.   But if we suppress  the tachyon
then this has already been accomplished in  (\ref{basiceq}) where
it is written as $ 2\ {\rm Im} <\Pi F, \Pi  G> $  .  Furthermore it 
is this choice  which is associated with positive energy as we shall see.
\end{enumerate}

These considerations lead  to the following definition.  Tachyons are 
completely suppressed. 
We bypass  $\si(\Phi, U)$ and go directly to operators 
$\Phi(F)$ satisfying (\ref{basic}).
Also we   enlarge the  class of test functions from compact support  
to the  Schwartz space 
of smooth rapidly decreasing functions.
The Hilbert space  is  the Fock space over the no-tachyon single string Hilbert space
$\cH_+$:
\begin{equation}
\cK = \cF(\cH_+)
\end{equation}
This has the indefinite inner product   $< \Psi, \Xi> = (\Psi, \Ga(\cJ) \Xi)$.
States in $\cK$ with finitely many entries are denoted  $\cK_f$.

\begin{defn}  (The String Field).
For   $  F \in \cS(\bbR^d  ,  \cF_+(\ell^2))$  we have   $\Pi F \in \cH_+$ and 
we define on    $\cK_f$   
\begin{equation}
\Phi (F)   =    a^{\dagger} (\Pi F) + a (\Pi F )
\end {equation}
\end{defn}

\begin{thm}  \
\begin{enumerate}
\item The string field  satisfies  $\Phi(F)^{\da} = \Phi(F)$, the field 
equation  $\Phi((-\square + M^2) F) =0$, and has the commutator 
$[ \Phi (F), \Phi(G)]  = -i  < F,EG >$.  
\item   There is positive energy  representation $\cU(a, \La)$
of the inhomogeneous Lorentz group on  $\cK$ such that 
\begin{equation}
\cU(a, \La)\Phi(F) \cU(a, \La)^{-1}
= \Phi (F_{a, \La}) 
\end{equation}
where  $F_{a, \La}(x)  =  \Ga(\La) F(\La^{-1}(x - a))$.
\end{enumerate}
\end{thm}
 
\pr   The field equation is satisfied since   $\Pi (-\square  + M^2 )F=0$.
The commutator is evaluated as  
\begin{equation}
[\Phi(F), \Phi(G) ]=  2i\ {\rm Im} <\Pi F, \Pi  G>   =  -i <F,EG>   
\end{equation}
since the identity  (\ref{basiceq})  holds for $\cS$ as well as  $\cC^{\infty}_0$.
The representation is defined by  $\cU(a, \La) =  \Ga ( U(a,\La))$.  
We compute 
\begin{equation}
\begin{split}
\cU(a, \La)\Phi(F) \cU(a, \La)^{-1}
=  & a^{\dagger} (U(a, \La)\Pi F) + a (U(a, \La)\Pi F )\\
=  & a^{\dagger} (\Pi F_{a, \La}) + a (\Pi F_{a, \La})\\
= & \Phi (F_{a, \La}) \\
\end{split}
\end{equation}

As noted we have the following corollary:

\begin{cor} ( Locality). If  $F,G$ have spacelike separated supports
$[ \Phi (F), \Phi(G)]  = 0$.  \label{vanishing}
\end{cor}
\bigskip

Now we impose the constraint, and just as for the single string
this will give us a positive definite inner product.  
Let $\hat L_m$ be the Fourier transform  of $L_m$, that is 
$\hat  L_m$ is given by  (\ref{L}) but with   $p_{\mu} = -i \pa / \pa x^{\mu}$.
We would like to select  states which are annihilated by 
$ L_m \Phi $ for  $m>0$.  However, just as  for 
the Gupta- Beuler quantization of the electromagnetic
field \cite{StWi74}
we must compromise and only impose the condition on the 
negative frequency  part of the field defined by 
$\Phi_{-} (F) =  a(\Pi F)$.  This is defined  and 
anti-linear on complex test functions.   We look for 
states annihilated by   $(\hat  L_m\Phi_-)(F) \equiv \Phi_-(  \hat L_{-m}F)$. 
This is fulfilled by taking the subspace  
\begin{equation}
\cK' =  \cF( \cH'_+) \ \ \ \ \ \ \ \ \ \ \ \ \ \ \ \ \  \cH'_+ = \cH' \cap \cH_+
\end{equation}
  All $\Psi \in \cK'$  satisfy the required  $a(\Pi \hat L_{-m} F)\Psi =0$
since if    $\psi \in \cH'_+$ then 
\begin{equation}
< \Pi \hat L_{-m} F, \psi>  = <  L_{-m} \Pi F, \psi>  =  <  \Pi F, L_m\psi> =0
\end{equation}
Thus  $\cK'$ is our constrained space,   something we might have guessed
directly.

Next  let   $\cK''  = \cK '  \cap (\cK')^{\perp}$ be the isotropic vectors in  $\cK'$.
and  define  
\begin{equation}
\cK^{phys} =   \cK'/ \cK''
\end{equation}
This space  inherits an indefinite inner product from  $\cK'$.

\begin{lem}  For $d=26$
the inner product on $\cK^{phys}$ is positive definite and we 
have 
the identification  of Hilbert spaces: 
\begin{equation}
\cK^{phys} =\cF(\cH^{phys}_+) \ \ \ \ \ \ \ \ \ \ \ \ \ \ \ \ \  \cH^{phys}_+= \cH'_+/\cH''_+
\end{equation}
\end{lem}

\pr    $\cH''_+ = \cH'' \cap \cH_+$ is a closed subspace of  $\cH'_+$ and so we can 
write   $\cH'_+ =  \cH''_+ \oplus \cM $ where   $\cM$ is the orthogonal 
complement with respect to the definite inner product. 
The projection onto   $\cM$   has kernel $\cH''_+$ and   gives
an identification of    $\cM$ with   $\cH^{phys}_+$ which preserves 
the indefinite inner product.  Thus the inner product is 
positive definite on  $\cM$, and of course zero if
either entry is in  $\cH''_+$.

Now we have  the identification of Hilbert spaces    \cite{GlJa71}  
\begin{equation}
\cK' =  \cF(\cH''_+ \oplus \cM ) =   \cF(\cH''_+) \otimes  \cF(\cM)
 \end{equation}
Under this identification $\Ga(\cJ)  =  \Ga(\cJ) \otimes \Ga(\cJ) $
and so   the induced indefinite inner product satisfies
$< \Psi_1 \otimes \Psi_2, \Psi'_1 \otimes \Psi'_2> =
< \Psi_1, \Psi'_1><  \Psi_2,  \Psi'_2> $.
Splitting  $ \cF(\cH''_+)=  \cF_0(\cH''_+)\oplus  \cF_+(\cH''_+)$
and using   $\cF_0(\cH''_+) \approx \bbC$ we have  
\begin{equation}
\cK'  =  \cF(\cM) \   \oplus ( \cF_+(\cH''_+) \otimes  \cF(\cM))
\end{equation}
with the natural indefinite inner product.  Every component 
of  $\cF_+(\cH''_+) \otimes \cF(\cM)$ has at least one factor 
in  $\cH''_+$ and so we can identify  
\begin{equation}
\cK'' =    \cF_+(\cH''_+) \otimes  \cF(\cM)
\end{equation} 
Thus      
\begin{equation}
\cK^{phys}  =    \cF(\cM)   =   \cF(\cH^{phys}_+)
\end{equation}
These identifications preserve the indefinite inner product.
Since the inner product is positive definite on  $\cH^{phys}_+$
it is positive definite on  $\cK^{phys} $.
\bigskip

For  certain test functions the 
string field operator  $\Phi(F)$   on $\cK$ determines an 
operator on  $\cK^{phys}$.  We define

\begin{defn}  
$F \in  \cS(\bbR^d, \cF_+(\ell^2))$
is a \emph{constrained} test function if   $\Pi F \in   \cH'_+$
\end{defn}

To get real  constrained test functions it is useful to pick 
a particular conjugation on $\cF(\ell^2)$.
It is  $C_1 = \Ga(c_1)$
where  $c_1$ on  $\ell^2$ is defined by 
$(c_1 f)_{0n} = \bar f_{0n}$ and  $(c_1 f)_{kn} = -\bar f_{kn}$ 
for $k=1,...,d-1$.    For the next result  real means  $C_1 \psi = \psi $.

\begin{lem} 
  Non-trivial  (real)  constrained test functions exist
\end{lem}

\pr    Take  $r \geq 0$ and choose $\psi_0 \in \cC^{\infty}_0 (V_r^+,\cF(\ell^2))$
so that   $\psi_0(p) \in \cH'(p)$ for all $p \in V_r^+$.  Then $\psi$ defines an element of
 $\cH_r'$ and hence an element of   $\cH'_+$.
We have seen that such  functions exist in lemma \ref{states}. 

Next define
\begin{equation}
\psi( \om_r(\vec p), \vec p) =  \psi_0( \om_r(\vec p), \vec p)
+C_1 \psi_0( \om_r(\vec p), -\vec p)
\end{equation}
This satisfies $ C_1\psi( \om_r(\vec p), \vec p)=\psi( \om_r(\vec p), -\vec p)$
and is still an element of   $\cH'_r$.   This is so since 
$C_1 \al^0_n C_1 = \al^0_n$  and  $C_1 \al^k_n C_1 = -\al^k_n$
and hence  $C_1 L_m( \om_r(\vec p), \vec p) C_1 =   L_m( \om_r(\vec p),- \vec p)$.

 We will  find  $F$ so that   $\Pi  F= \psi$.
First write  $\psi(\om_r(\vec p), \vec p) = h(\vec p)$ for a  function
$h \in \cC^{\infty}_0( \bbR^{d-1}, \cF(\ell^2))$. (Or $\bbR^{d-1} - \{ 0\}$ if  $r=0$).
Let $\chi \in \cC^{\infty}_0 (\bbR)$ be real and satisfy    $\chi(0) =1$. 
 We define $F$ by specifying that the Fourier transform  be
\begin{equation}
\tilde F (p^0, \vec p)  = (2 \pi)^{-(1/2) } \chi(-(p^0)^2 + |\vec p|^2 +r )h( \vec p) 
\end{equation}
 Then $\tilde F$  is smooth and has compact support and
 hence   $F \in  \cS(\bbR^d, \cF(\ell^2))$.  Since  $h( \vec p)  \in Ran
P_r$
 we have $(\Pi F)_r = \psi$ and    $(\Pi F)_{r'} = 0$ for $r' \neq r$ as required.
Since elements of $Ran P_r$ have no zero component in Fock space,  this is true  
of $h( \vec p) $ and hence  $F(x)$.  Thus  $F$ takes values
in  $\cF_{+}$.    Finally we have    $C_1 \tilde F (p)  =  \tilde F (-p)$
and hence   $C_1 F(x)= F(x)$  so  $F$ is real.
\bigskip

Recall that     $\cK_{f}$  is the subspace of  $\cK$ with a 
finite number of entries.  Similarly define  $\cK'_f$
and $\cK''_f$  and  $\cK^{phys}_f  =
\cK'_f/\cK''_f$.  One can identify    $\cK^{phys}_f$  
 with a dense subspace of  $\cK^{phys}$.

\begin{thm}  (Observable fields for $d=26$)
\begin{enumerate}
\item 
Let    $F  \in   \cS(\bbR^d, \cF_+(\ell^2))$  be a constrained test function. Then 
$\Phi(F)$ on $\cK_f$ lifts to  an operator  $\Phi(F) $ on  $\cK^{phys}_f$ called
an observable field
\item    These satisfy  
$[ \Phi (F), \Phi(G)]  = -i  < F,EG >$.  
\item 
 The  representation $\cU(a, \La)$ on  $\cK$ lifts to a unitary  representation
$\cU(a, \La)$ on  $\cK^{phys}$ and 
 \[\cU(a, \La)\Phi(F) \cU(a, \La)^{-1}
= \Phi (F_{a, \La}) \]
\end{enumerate}
\end{thm}

 \pr Since 
    $\Pi F \in  \cH'_+$
we have that    $\Phi(F)$ preserves   $\cK'_f$.
It also preserves  $\cK''_f$ since if $\Psi  \in \cK''_f$
and     $\Xi \in \cK_f'$
then 
$< \Xi,  \Phi(F)  \Psi>=  < \Phi(F)\Xi,   \Psi>=0$.
Since   $\cK'_f$ is dense in $\cK'$
 we have  $< \Xi,  \Phi(F)  \Psi>=0$ for all $\Xi \in \cK'$ 
and hence  $ \Phi(F)  \Psi \in \cK''_f$.
Hence   $\Phi(F)$ acts on  $\cK^{phys}_f$.

The commutator  follows from the commutator on  $\cK_f$

For the covariance first note that  $F$ is constrained if and only if  $F_{a, \La}$
is constrained.
This  follows from the identity
$\Pi F_{a, \La} = U(a, \La) \Pi F$ and the fact that  $U(a, \La)$
preserves  $\cH'_+$.
The operator  $\cU(a, \La)$ preserves  $\cK'_f$ since 
$U(a, \La)$ preserves  $\cH'_+$.  We argue as before that
it also preserves  $\cK''_f$ and so it lifts.  
The unitarity follows since it is inner product preserving,
and the identity lifts from the identity on  $\cK_f$.
\bigskip

\re   According to this theorem the observable fields
have  a local commutator.
But can the fields themselves be localized? That is, are
there constrained
test functions   $F  \in   \cC^{\infty}_0(\bbR^d, \cF_+(\ell^2))$? 
Or is there some other way to get strictly localized operators?
These are open questions.
Without strictly localized fields we cannot get a 
vanishing result like   Corollary  \ref{vanishing}.  The best we 
can do is the following approximate result.  If $F,G$ are 
constrained then it estimates the commutator for observable 
fields on $\cK^{phys}$.  Otherwise it refers to fields on  $\cK$.

\begin{cor} let   $F,G   \in   \cS(\bbR^d, \cF_+(\ell^2))$,
and let   $a$ be in the  spacelike   region   $|a^0|  < (1-\ep)|\vec a|$. 
Then as  $|a| \to \infty$, we have for any  $n>0$
\begin{equation}
[\Phi(F_a), \Phi(G)]   = -i <F_a, EG>  =\cO( |a|^{-n} )
\end{equation}
\end{cor}

\pr Since    $EG$
is bounded we have  
\begin{equation}
\begin{split}
|<F_a, EG>|  \leq &\one \int_{supp(EG)} \|F(x-a)\| dx\\
 \leq & \one \int_{supp(EG)} (1 + |x-a|)^{-n-d-1} dx\\
\leq &   \one  d(a,  supp (EG))^{-n} \\
\leq &   \cO ( |a|^{-n}) \\
\end{split}
\end{equation}
In the last step we  use the fact that  $supp (EG)$ is contained 
in a set of the form  $\{ x \in  \bbR^d:  |\vec x| \leq x^0 +C \}$.
We omit the details.

\end{document}